\documentclass[sigconf,anonymous=false]{acmart}

\setcopyright{rightsretained}

\AtBeginDocument{%
  \providecommand\BibTeX{{%
    \normalfont B\kern-0.5em{\scshape i\kern-0.25em b}\kern-0.8em\TeX}}}

\copyrightyear{2021}
\acmYear{2021}

\acmConference[KDD '21]{Knowledge Discovery and Data Mining}{August 14--18, 2021}{Virtual Event, Singapore}
\acmBooktitle{ACM KDD AI4Cyber: The 1st Workshop on Artificial Intelligence- enabled Cybersecurity Analytics at KDD'21, August 14--18, 2021, Virtual}

\newcommand{\bron}{\textit{BRON}\xspace}
\newcommand{\tcedge}{Technique-CAPEC\xspace}
\newcommand{\Attack}{ATT\&CK\xspace}
\newcommand{\TACTICS}{\textit{Tactics}\xspace}  
\newcommand{\TECHNIQ}{\textit{Technique}\xspace}  
\newcommand{\TECHNIQS}{\textit{Techniques}\xspace}  
\newcommand{\WEAKNESS}{\textit{Weakness}\xspace}  
\newcommand{\VULNERABILITIES}{\textit{Vulnerabilities}\xspace} 

\newcommand{\shortconfig}{\textit{Affected Prod Conf}}
\newcommand{\config}{\textit{Affected Product Configuration}\xspace}  
\newcommand{\APC}{\textit{APC}\xspace}

\newcommand{\AttackPattern}{\textit{Attack Pattern}\xspace}

\newcommand{\ATTACKPATTERN}{\textit{Attack Pattern}\xspace}
\newcommand{\ATTACKPATTERNS}{\textit{Attack Patterns}\xspace}
\newcommand{\ap}{\ATTACKPATTERN}

\newcommand{\Inference}{Inference\xspace}

\usepackage{lipsum}
\usepackage{amsfonts}
\usepackage{graphicx}
\usepackage{epstopdf}
\usepackage{algorithmic}
\usepackage{longtable}
\usepackage{booktabs}
\usepackage{xspace}
\usepackage{multirow}
\usepackage{paralist}
\usepackage{tikz}
\usepackage{url}
\usepackage{subcaption}

\begin{document}

\title{Using a Collated Cybersecurity Dataset for Machine Learning and Artificial Intelligence}

\author{Erik Hemberg}
\affiliation{%
  \institution{MIT CSAIL}
  \streetaddress{32 Vassar Street}
  \city{Cambridge}
  \country{United States of America}}
\email{hembergerik@csail.mit.edu}

\author{Una-May O'Reilly}
\affiliation{%
  \institution{MIT CSAIL}
  \streetaddress{32 Vassar Street}
  \city{Cambridge}
  \country{United States of America}}
\email{unamay@csail.mit.edu}

\newcommand{\capec}{CAPEC\xspace}
\newcommand{\cve}{CVE\xspace}
\newcommand{\cwe}{CWE\xspace}


\begin{abstract}
Artificial Intelligence~(AI) and Machine Learning~(ML) algorithms can support the span of indicator-level, e.g. anomaly detection, to behavioral level cyber security modeling and inference.  This contribution is based on a dataset named \bron which is amalgamated from public threat and vulnerability behavioral sources. We demonstrate how \bron can support prediction of related threat techniques and attack patterns. We also discuss other AI and ML uses of \bron to exploit its behavioral knowledge.
\end{abstract}

\ccsdesc[500]{Security and privacy}
\ccsdesc[500]{Computing methodologies~Artificial intelligence}
\ccsdesc[500]{Computing methodologies~Machine learning}
\ccsdesc[500]{Computing methodologies~Machine learning approaches}

\keywords{cyber security, threat hunting, Machine Learning, prediction}

\maketitle

\section{Introduction}
\begin{figure}[tb]
\includegraphics[width=0.49\textwidth]{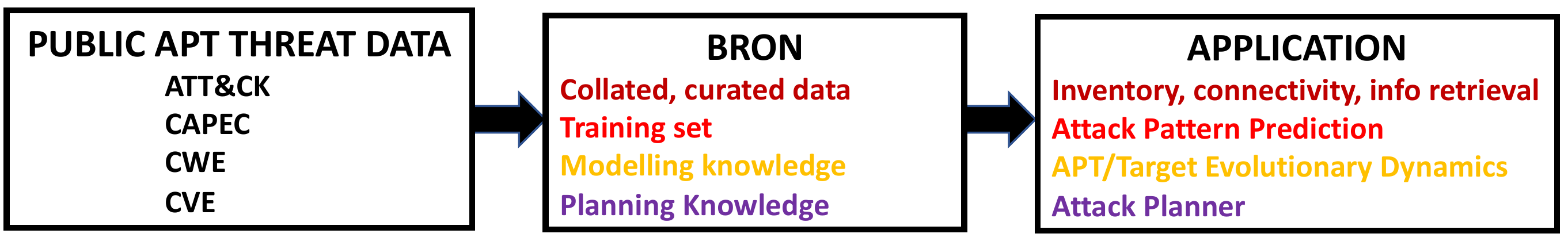} 
\centering
\caption{High level concept of \bron. Shows the data sources used and possible applications.}
\label{fig:bron-highlevel}
\end{figure}

Among other enticements, Artificial Intelligence~(AI) and Machine
Learning~(ML) offer attack planning, defensive modeling, threat prediction, anomaly detection, and simulation of adversarial
dynamics in support of
cyber security~\cite{falco2018master,ghafir2018detection,husari2017ttpdrill,araujo2021evidential,dhir2021prospective,elitzur2019attack,afzaliseresht2020investigating}.
Automated security activity presently mainly focuses on lower level malicious activity detection that relies upon indicators of compromise and forensics.
Alternatively, AI and ML techniques for cyber that work at the behavioral level are emerging. They  typically draw upon threat  information that abstractly describes an attacker's tactics, techniques and procedures (TTPs) as well as vulnerability knowledge such as exposed product configurations and system weaknesses. These information sources are typically independent, though they sometimes have external links to one another.
 Here we demonstrate the use of a single dataset, \bron\footnote{\bron means bridge in Swedish, referring to how it links data sources}, that supports AI modeling and ML inference at a
behavioral level, see Figure~\ref{fig:bron-highlevel}, by using an amalgamated set of key public threat and vulnerability information sources. \bron is fully described in \cite{hemberg2020bron}.  

Public threat and vulnerability information is, unfortunately, extracted
from historic attacks, such as Advanced Persistent
Threats (APTs). Post-hoc, APTs are catalogued and framed as the behavior of a
specific actor pursuing a goal, posing a threat that has specific
tactics, techniques and procedures. The targets of attacks are
itemized as hardware or software vulnerabilities or exposures which
are sometimes themselves cross-referenced to a type of weakness as
found in code, design, or system architecture. Attack patterns are
recognized manually and enumerated. According to its type, each unit
of information is populated as an entry of a specific database, with
some amount of cross-referencing. The combined databases, with
irregular, pairwise linkages between them, serve defensive reasoning.

This contribution demonstrates the use of the combined data of
four such public databases amalgamated into a single graph database,
\bron\cite{hemberg2020bron}. The four collated databases are:
\begin{asparaenum}[\itshape a)]
\item MITRE’s ATT\&CK MATRIX of \TACTICS, \TECHNIQS, Procedures and Sub-techniques~\cite{attack}
\item MITRE’s Common Attack Pattern Enumeration and Classification dictionary~(CAPEC)~\cite{capec}
\item MITRE’s Common Weakness Enumerations~(CWE)~\cite{cwe}
\item NIST’s Common Vulnerabilities and Exposures~(CVE)~\cite{nvd}
\end{asparaenum}
Their collective entries, and links between entries, are stored in
\bron, a threat and vulnerability graph database.  \bron adds no new
information, while it adds bi-directional links to enable faster and
more convenient queries. It is publicly available and regularly
updated at \url{http://bron.alfa.csail.mit.edu}.

The combined information on APTs within \bron expresses, for a threat, who is behind, how it works and what it targets. For a vulnerability, it expresses its type of
weakness, and how it can be threatened.  The structure of this
information allows \bron to support statistical ML and inference. We
illustrate this in Section~\ref{sec:ml} with the problem of predicting
edges that exist between entries, but which have not been
reported. Solving this pattern inference problem would benefit cyber
security experts in partially resolving the ambiguity of missing
edges; a trained predictive model could suggest probable edges.

We then, in Section~\ref{sec:discussion} discuss how \bron can be used for other benefits. We include information retrieval (Section~\ref{sec:inform-retr}) showing how \bron can be queried to inform users of the original sources or \bron about connections that are not present. Connections that are not present are ambiguous: Is the relational behavior non-existent, or existent but missing?  We include modeling and simulation (Section~\ref{sec:modsim})  describing two types of use and a coevolutionary simulation of APT threats and mitigations. We finally cover AI planning (Section~\ref{sec:ai}) where \bron functions as a knowledge base for attack planning or planning on
attack graphs.  This could, for example. assist with automation of red-teaming. Finally, future work directions  are presented in Section~\ref{sec:concl-future-work}.

%

\section{Pattern \Inference}
\label{sec:ml}

\bron can be a source of training data for machine
learning~\cite{brondb,brondbGitHub,bronML}. Graph properties such as
number of incoming or outgoing edges, or number of paths of different
connections and lengths can be used as features or labels. As well,
the natural language part of entries in \bron offers semantic
value that can be featurized.

One particularly challenging ML problem exists within \bron (and among
the independent databases) due to the irregular nature of the
cross-database linkages. Links only exist if they have been noted and
reported.  \bron, \textit{circa} May 2021, has 666 \TECHNIQS and 740
\ap{s}, so, in theory, there are a total of 492,840 possible
\TECHNIQ-\ap connections.  \bron reveals a total of 157 connected
\TECHNIQ-\ap connections, a percentage of merely 0.032\%.  This is
accurate, to the extent that most of the possible connections would
never be semantically sensible. However, given about 74\% of \TECHNIQS
are not linked to a \ap and the stealthy nature of APTs, there are
very likely connections that are not noted or undetected.  It also
follows then that the goal to predict links (edges) between ATT\&CK
\TECHNIQS and \ap{s} (nodes) is important but complex and one of
imputation.

%
%

\paragraph{Use-case: BRON for \tcedge edge prediction}

Th goal is to predict links (edges) between ATT\&CK Techniques and
CAPECs (nodes). Each node has a textual name (e.g. “Interception”).
As an initial study we use this textual information to predict \tcedge
edges.

We are interested in:
\begin{inparaenum}
\item the difference in performance due to feature selection, i.e. when
  more data is used from BRON,
\item the change in performance due to the feature representation,
\item a baseline classification performance established from untuned
  classifier models.
\end{inparaenum}
Thus, we formulate a supervised binary classification problem: given
information on pairs of one Technique and one CAPEC entry, train an
inference model that predicts if there is a link between the pair of
entries. 


We encode the text information on Techniques and CAPECs using natural language
processing into a vector as the input to the model.
 The entity names of different \bron data sources as \textit{feature selection (data sources)} we use combinations of:
 \begin{inparaenum}[\itshape a)]
 \item CAPEC 
   \item Techniques
 \item Tactics
 \item CWE
 \item CAPEC\_Techniques, refers to the names of all known
Techniques connected to the CAPEC but not including the name of the
CAPEC itself.
\end{inparaenum}
 For example, we create a string for each of the selected features of
 Tactic, Technique, CAPEC and CWE: \textit{Discovery, System Network
   Configuration Discovery, Network Topology Mapping, Exposure of
   Sensitive Information to an Unauthorized Actor}. Each string is
 encoded according to some \textit{feature representation}. We experiment with two different feature
 representations:
\begin{inparaenum}[\itshape a)]
\item Bag-of-Words (BOW),
\item Transformer Neural Network, BERT~\cite{devlin2018bert}.
\end{inparaenum}
Seven different \textit{Classification methods} with default settings
from SciKit-Learn~\cite{scikit-learn}:
\begin{inparaenum}[\itshape a)]
\item Multi Layer Perceptron (MLP),
\item Random Forest,
\item Logistic Regression (SGD),
\item K-Nearest Neighbor (KNN).
\item Naive Bayes (NB),
\item Support Vector Machine, linear kernel (SVM)
\item Support Vector Machine, radial basis function kernel (RBF-SVM)
\end{inparaenum}
In total we have 84 different combinations of features
selections, features representations and classification methods.
 
For all combinations we measure \textit{performance} with:
 \begin{inparaenum}[\itshape a)]
 \item Error (1.0 - accuracy),
 \item AUC,
 \item F1-score.
\end{inparaenum}
We perform 100 independent trials with different 70-30 train-test data
splits. The data for each trial has 314 exemplars with 50-50 class
balance (from under-sampling the majority class).

\paragraph{Results \& Discussion}

\begin{figure}[!tb]
  \centering
  \begin{subfigure}{0.43\textwidth}  
    \centering
    \includegraphics[width=0.99\textwidth]{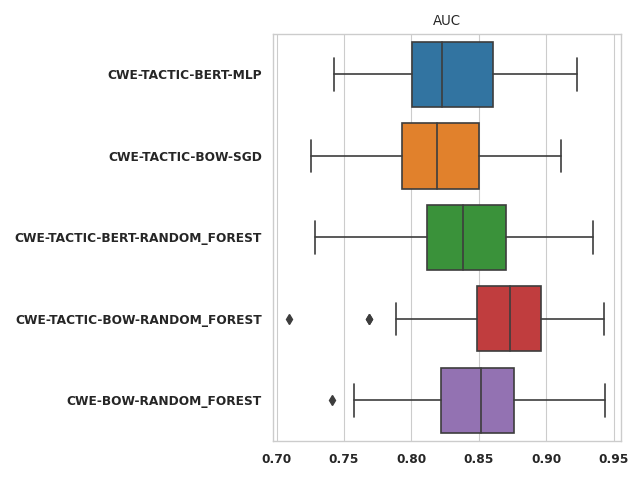} 
    \caption{AUC, higher is better}
    \label{fig:bron_link_prediction_results_auc}
    \end{subfigure}
  \begin{subfigure}{0.43\textwidth}  
    \centering
    \includegraphics[width=0.99\textwidth]{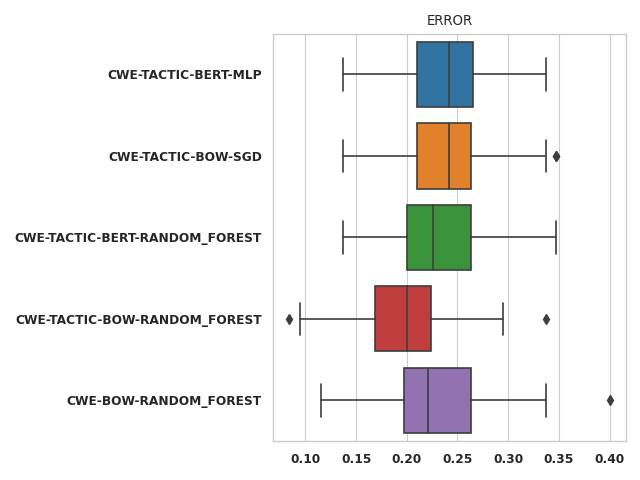} 
    \caption{Error, lower is better}
    \label{fig:bron_link_prediction_results_error}
    \end{subfigure}
  \begin{subfigure}{0.43\textwidth}  
    \centering
    \includegraphics[width=0.99\textwidth]{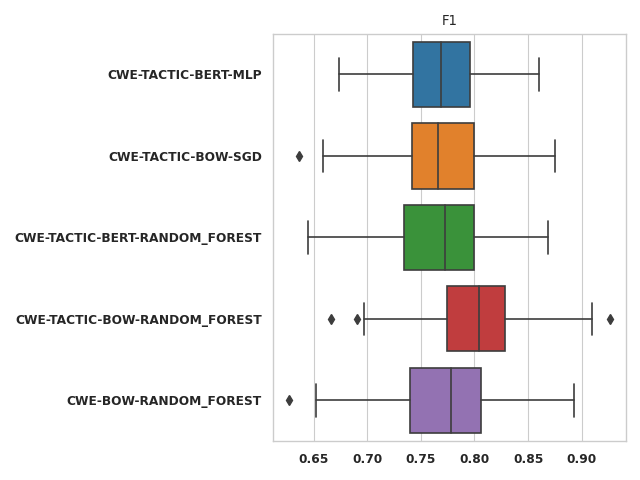} 
    \caption{F1, higher is better}
    \label{fig:bron_link_prediction_results_f1}
  \end{subfigure}
  \caption{Box-plots of the performance measures for different \tcedge
    link predictions. The experiment name indicates what data sources,
    features and classifier was used (- is a separator of features,
    representation and classifier).}
\label{fig:bron_link_prediction_results}
\end{figure}

Figure~\ref{fig:bron_link_prediction_results} and
Table~\ref{tab:results} show results from predicting \tcedge with data
from \bron. We see that using data from \bron can improve the
performance. The experiment name indicates what data sources, features
and classifier was used. For readability and space considerations we
only show the top 5 based on F1-score. The significance tests reveal
that CWE-TACTIC-BOW-RANDOM\_FOREST has better performance on each of
the measures (Error, AUC and F1). We measure if the differences in
mean are statistically significant with a Wilcoxon-ranksum test,
Bonferroni post-hoc adjustment and a p-value threshold of 0.05.

\begin{table}[tb!]
  \footnotesize
  \centering
  \caption{Mean performance measures for top 5 experiments. Bold
    indicates the best value.}
  \label{tab:results}
  \begin{tabular}{l|lll}
\textbf{Name} & \textbf{Error} & \textbf{AUC} & \textbf{F1} \\
\hline
CWE-TACTIC-BOW-SGD & 0.238 & 0.821 & 0.768 \\
CWE-TACTIC-BERT-MLP & 0.242 & 0.827 &  0.769 \\
CWE-TACTIC-BERT-RANDOM\_FOREST &  0.231 &  0.840 &  0.770 \\
CWE-BOW-RANDOM\_FOREST & 0.226 & 0.850 & 0.773 \\
CWE-TACTIC-BOW-RANDOM\_FOREST &  \textbf{0.197} &  \textbf{0.870 } & \textbf{ 0.802} \\
  \end{tabular}
\end{table}

The feature selection experiments showed that in general more data
improved performance, however the CAPEC\_TECHNIQUE feature was not
used by any of the top five experiments. In regard to feature
representation BOW seem to work well, however BERT might improve with
more cybersecurity specific training vocabulary and more data.  The
best classifiers were Random Forest.  We see that using linked
data sets from \bron can improve the performance. As expected, there is
a difference between default classifier performance, and performance
can hopefully be improved with parameter tuning. These experiments can
be extended with more or other data.






\section{Discussion}\label{sec:discussion}

In this section we discuss how \bron can be used for information
retrieval, modeling and simulation, and AI planning.

\subsection{Information Retrieval}
\label{sec:inform-retr}

\paragraph{Example: Analyzing top 25 CWEs with \bron}

The \textit{2020 Common Weakness Enumeration (CWE) Top 25 Most
  Dangerous Software Weaknesses} list~\cite{topcwe} is compiled by
considering the prevalence and severity of CVEs and their associated
\WEAKNESS{es} (implemented by linking). These \WEAKNESS{es} highlight
``the most frequent and critical errors that can lead to serious
vulnerabilities in software''~\cite{topcwe}. For example, an attacker
can exploit the vulnerabilities to take control of a system, obtain
sensitive information, or cause a denial-of-service.  The CWE Top 25
list is a resource that can provide insight into the most severe and
current security weaknesses~\cite{topcwe}.

We use \bron to answer the following questions. What are the \TACTICS,
\TECHNIQ{s} and \ATTACKPATTERNS linked to these \WEAKNESS{es}? What
commonalities in these features are there among the Top 25?

\begin{table*}[tb]
\footnotesize
\centering
\caption{Top 25 CWE~\cite{topcwe}. \APC is \config.}
\label{tab:topcwe}
\begin{tabular}{lp{5cm}lllllll}
\textbf{CWE ID} & \textbf{Name} & \textbf{\#\TACTICS} & \textbf{\#\TECHNIQS} &
\textbf{ \#\ATTACKPATTERNS} & \textbf{ \#\VULNERABILITIES} & \textbf{Sum CVSS} & \textbf{Ave CVSS} & \textbf{ \#\APC}\\
\hline

CWE-79
&
{Improper Neutralization of Input During Web Page Generation
('Cross-site Scripting')}
&
{0}
&
{0}
&
{6}
&
{12,629}
&
{57,890}
&
{4.58}
&
{43,013}
\\

CWE-787
&
{Out-of-bounds Write}
&
{0}
&
{0}
&
{0}

&
{1,159}
&
{8,667}
&
{7.48}
&
{1,499}
\\

CWE-20
&
{Improper Input Validation}
&
{1}
&
{3}
&
{51}
&
{7,820}
&
{49,995}
&
{6.39}
&
{44,399}
\\

CWE-125
&
{Out-of-bounds Read}
&
{0}
&
{0}
&
{2}
&
{2,172}
&
{13,549}
&
{6.24}
&
{2,029}
\\

CWE-119
&
{Improper Restriction of Operations within the Bounds of a Memory
Buffer}
&
{0}
&
{0}
&
{12}
&
{10,449}
&
{81,125}
&
{7.76}
&
{39,673}
\\

CWE-89
&
{Improper Neutralization of Special Elements used in an SQL Command
('SQL Injection')}
&
{0}
&
{0}
&
{6}
&
{5,477}
&
{41,309}
&
{7.54}
&
{14,685}
\\

CWE-200
&
{Exposure of Sensitive Information to an Unauthorized Actor}
&
{2}
&
{14}
&
{58}
&
{6,824}
&
{32,813}
&
{4.81}
&
{29,874}
\\

CWE-416
&
{Use After Free}
&
{0}
&
{0}
&
{0}
&
{1,187}
&
{8,742}
&
{7.37}
&
{1,636}
\\

CWE-352
&
{Cross-Site Request Forgery (CSRF)}
&
{0}
&
{0}
&
{4}
&
{2,377}
&
{16,319}
&
{6.87}
&
{11,646}
\\

CWE-78
&
{Improper Neutralization of Special Elements used in an OS Command ('OS
Command Injection')}
&
{0}
&
{0}
&
{5}
&
{724}
&
{6,131}
&
{8.47}
&
{3,325}
\\

CWE-190
&
{Integer Overflow or Wraparound}
&
{0}
&
{0}
&
{1}
&
{1,218}
&
{8,268}
&
{6.79}
&
{2,129}
\\

CWE-22
&
{Improper Limitation of a Pathname to a Restricted Directory ('Path
Traversal')}
&
{0}
&
{0}
&
{5}
&
{2,964}
&
{18,684}
&
{6.3}
&
{14,368}
\\

CWE-476
&
{NULL Pointer Dereference}
&
{0}
&
{0}
&
{0}
&
{1,019}
&
{5,994}
&
{5.88}
&
{2,342}
\\

CWE-287
&
{Improper Authentication}
&
{4}
&
{3}
&
{10}
&
{1,654}
&
{11,453}
&
{6.92}
&
{13,061}
\\

CWE-434
&
{Unrestricted Upload of File with Dangerous Type}
&
{0}
&
{0}
&
{1}
&
{562}
&
{4,315}
&
{7.68}
&
{1,370}
\\

CWE-732
&
{Incorrect Permission Assignment for Critical Resource}
&
{0}
&
{1}
&
{11}
&
{427}
&
{2,654}
&
{6.22}
&
{1,151}
\\

CWE-94
&
{Improper Control of Generation of Code ('Code Injection')}
&
{0}
&
{0}
&
{3}
&
{2,287}
&
{17,683}
&
{7.73}
&
{12,666}
\\

CWE-522
&
{Insufficiently Protected Credentials}
&
{5}
&
{15}
&
{9}
&
{277}
&
{1,548}
&
{5.59}
&
{923}
\\

CWE-611
&
{Improper Restriction of XML External Entity Reference}
&
{0}
&
{0}
&
{1}
&
{488}
&
{3,381}
&
{6.93}
&
{1,985}
\\

CWE-798
&
{Use of Hard-coded Credentials}
&
{0}
&
{0}
&
{2}
&
{244}
&
{1,919}
&
{7.87}
&
{543}
\\

CWE-502
&
{Deserialization of Untrusted Data}
&
{0}
&
{0}
&
{1}
&
{387}
&
{3,151}
&
{8.14}
&
{1,580}
\\

CWE-269
&
{Improper Privilege Management}
&
{0}
&
{0}
&
{3}
&
{1,095}
&
{7,421}
&
{6.78}
&
{3,770}
\\

CWE-400
&
{Uncontrolled Resource Consumption}
&
{0}
&
{0}
&
{3}
&
{728}
&
{4,459}
&
{6.13}
&
{4,303}
\\

CWE-306
&
{Missing Authentication for Critical Function}
&
{0}
&
{0}
&
{4}
&
{112}
&
{793}
&
{7.09}
&
{504}
\\

CWE-862
&
{Missing Authorization}
&
{0}
&
{0}
&
{0}
&
{190}
&
{1,162}
&
{6.12}
&
{527}
\\
\end{tabular}
\end{table*}


Our analysis of the Top 25 CWE~\cite{topcwe} is summarized in
Table~\ref{tab:topcwe}. We observe 4 of 25 \WEAKNESS{es} lack a
presence in any \AttackPattern. Reflecting diversity in threats that
could target the \WEAKNESS{es}, there are 8 distinct \TACTICS
associated with the Top 25. In terms of commonalities, the most
frequent \TACTICS associated with the Top 25 weakness{es} are
\texttt{Defense Evasion, Privilege escalation, Discovery}. Only 2
\TECHNIQ{s}, \texttt{T1148, T1562.003} occur more than once. The three
most frequent \ATTACKPATTERNS are \texttt{Using Slashes in Alternate
  Encoding, Exploiting Trust in Client, Command Line Execution through
  SQL Injection}.

The three most frequent \VULNERABILITIES (i.e. top 3) occur 3 times
each and are \texttt{CVE-2017-7778, CVE-2016-10164,
  CVE-2016-7163}. The three most frequent \shortconfig(s) are 3
different linux versions occurring 23, 24, 25 times respectively.  We
also analyzed the \WEAKNESS text descriptions with a frequency
analysis of unigrams and bigrams. \texttt{Buffer Overflow} emerged as
a common \ATTACKPATTERN bigram.

We note that not all weakness are linked with the same frequency to
\ATTACKPATTERN{s}, \TECHNIQS and \TACTICS.  The ambiguity of this
finding prompts: is absent data due to non-existence or being
unreported? In addition, each of the source datasets has some
bias. E.g. ATT\&CK is from APT groups and include only common tactics
and techniques.  CWE and CVE include unexploited software
vulnerabilities.  BRON can help make sense of connected data before
using them for AI/ML.

\subsection{Modeling and Simulation}
\label{sec:modsim}

\bron can also be used for modeling and simulation (ModSim). Related
works in cybersecurity use ModSim to conduct sensitivity analysis of
network vulnerabilities and threats, or to investigate dynamics
between threat and defense adaptations. The latter class of works
intersects with studies of the coevolution of attacks and
defenses\cite{o2020adversarial}. While it does not use \bron,
\cite{kelly2019adversarially} models the reconnaissance stage behavior
of an APT and the deceptive cloaking of a software-defined network. It
simulates the behaviors coevolving through using feedback to adapt
after engagements where a reconnaissance scan tries to operates within
a defensive overlay.

A \bron-based example is EvoAPT\cite{shlapentokhRothman2021}. This
evolutionary algorithm system incorporates known threats and
vulnerabilities from \bron into a stylized "competition'' that pits
cyber \ap{s} against \textit{mitigations}. The outcome of a
competition is quantified using the Common Vulnerability Scoring
System - CVSS, values within \bron. Variations of \ap{s} within the
simulation are drawn from \bron. Mitigations take two forms: software
updates or monitoring, and the software that is mitigated is
identified by drawing from \bron's entries from the CVE database.
Three abstract models of population-level dynamics where APTs interact
with defenses are aligned with three competitive, coevolutionary
algorithm variants that use the competition.  A comparative study
shows that the way a defensive population preferentially acts,
e.g. shifting to mitigating recent attack patterns, results in
different evolutionary outcomes, expressed as different dominant
attack patterns and mitigations.

We foresee \bron supporting other ModSim environments and studies.  We
anticipate that it will be plumbed for its APT behavioral structure,
its connective structure, and text, offering further possible
elaboration of modeled APT behaviors.

\subsection{Planning}
\label{sec:ai}

\bron can be incorporated into traditional artificial intelligence
(AI) planning.  One use case is driven by a need for automated
red-teams which attack a system to gauge its defensive capacity or the
competence of its security team~\cite{reinstadler2021}.  The attacks
can be plans derived by planners. The planner, itself, requires
structured threat data and guidance on how to make domain-specific
adaptations.

One close example is ~\cite{reinstadler2021}. This system utilizes a
complex knowledge base which references APT information from \Attack.
Another example, that specifically incorporates \bron is, Attack
Planner~\cite{nguyen2021}.  It is a computational vulnerability
analysis system that outputs multistage attack model
trees that achieve a desired goal on a desired system resource. Attack
Planner generates attack graphs to achieve different goals, based on
already known tactics and techniques. In order to incorporate \Attack
and CVE, \bron was used via an interface between \bron’s graph
representation of this data and the Attack Planner. \Attack and CVE
categorize and organize all stages of an attack campaign at varying
levels of depth starting from an overarching goal to down to specific
exploits on a specific version of an operating system. By using \bron
to link the specific exploits with their parent goals, the Attack
Planner is able to generate plans with higher detail.

\section{Summary and Future Work}
\label{sec:concl-future-work}
We have demonstrated and discussed how \bron, a collated information
dataset supports ML and AI at the behavioral level. Uses of \bron
include information retrieval, pattern inference, modeling and
simulation and AI-based attack planning.

The inference could be improved by tuning the feature representation
and classifiers. \bron could be enhanced with additional behavioral
knowledge, from timely sources such as threat reports.  It could also
be the basis of an open challenge within a security, knowledge
discovery or applied ML workshop. E.g. extend inference or formulate
more supervised learning problems around missing data.


\paragraph*{Acknowledgments}
This material is based upon work supported by the DARPA Advanced
Research Project Agency (DARPA) and Space and Naval Warfare Systems
Center, Pacific (SSC Pacific) under Contract No. N66001-18-C-4036

\newpage

\bibliographystyle{ACM-Reference-Format}
\bibliography{references}
\end{document}